\documentclass[onecolumn]{article}
\usepackage{graphicx}
\usepackage{epstopdf}
\usepackage{epsfig}
\usepackage{dcolumn}
\usepackage{amssymb,amsmath}
\usepackage{color}
\usepackage{soul}

\def\be{\begin{equation}}
\def\ee{\end{equation}}

\usepackage[font=small,labelfont=bf]{caption}

\begin{document}
\bibliographystyle{apsrev}
\title{Continuous hydraulic jumps in laminar channel flow}

\author{Dimitrios Razis\textsuperscript{1}, Giorgos Kanellopoulos\textsuperscript{1} and Ko van der Weele\textsuperscript{1}\\
\small{\textsuperscript{1} Department of Mathematics, University of Patras, 26500 Patras, Greece}}

\date{}

\maketitle


\begin{abstract}
\noindent On the basis of the viscous Saint-Venant equations, hydraulic jumps in laminar open channel flow are obtained as \textit{continuous} shock structures. Thanks to the inclusion of viscosity, the jumps are not abrupt, rendering the classic patchwork via the Rankine-Hugoniot shock relations unnecessary. The jumps arise as stable stationary solutions of the governing equations and lend themselves excellently to a Dynamical Systems analysis, manifesting themselves as near-parabolic trajectories in phase space. Based on this, we derive an analytic expression for the jump length as a function of the Froude and Reynolds numbers, reflecting the fact that both gravity and viscosity contribute to the balance of forces that shape the jump. The paper concludes with a numerical experiment confirming the stability of the jumps.
\vspace{10pt}


\end{abstract}


\section{Introduction}
\label{sec:intro}

\begin{figure}
\begin{center}
\includegraphics[width=0.65\textwidth]{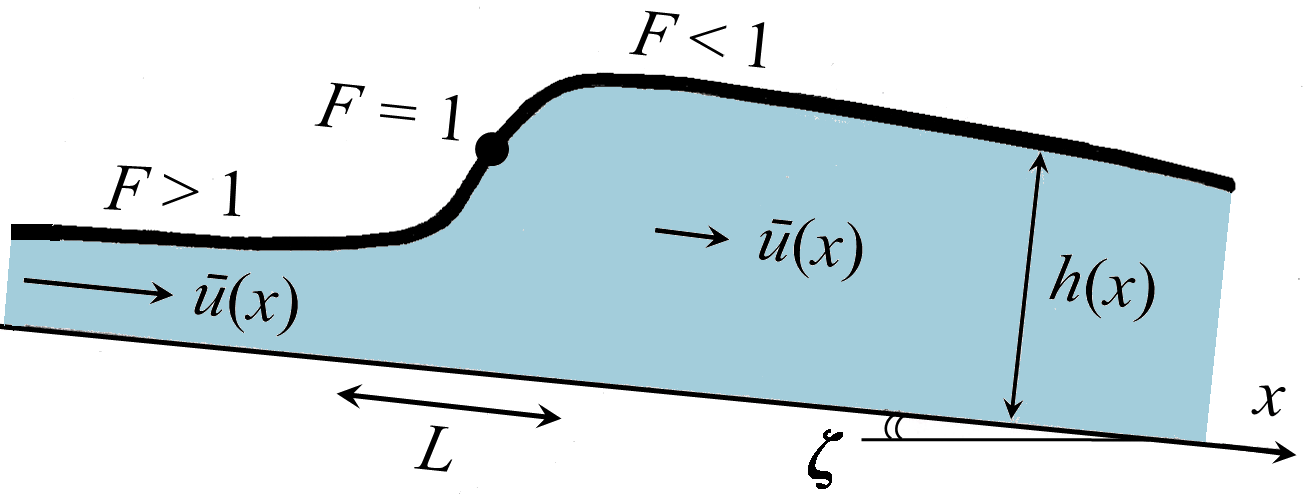}
\caption{Schematic view of a hydraulic jump in a mildly tilted channel, i.e., the transition zone of length $L$ from a shallow supercritical flow (Froude number $F>1$) to a thicker subcritical one ($F<1$). The flow is fully described by the height $h(x)$ and the depth-averaged velocity $\bar{u}(x)$, governed by the viscous Saint-Venant equations (\ref{massbal})-(\ref{mombal}). The depicted jump corresponds to the profile $\widetilde{M}3 \rightarrow \widetilde{M}2$ of Fig.~\ref{fig2}a, which is one of the four qualitatively different types of laminar jumps predicted by our analysis, cf. Fig.~\ref{fig3}.} \label{fig1}
\end{center}
\end{figure}

Hydraulic jumps are one of the most iconic phenomena in fluid dynamics, found in a variety of situations ranging from thin films and kitchen sinks to large scale open channels. They mark the sudden transition from a fast supercritical flow (thin) to a slower subcritical one (thick); see Fig.~\ref{fig1}. The systematic study of hydraulic jumps was initiated in the first half of the 19th century by the experiments of Bidone and the theoretical investigations of B\'elanger~\cite{bakhmeteff1932,chanson2009}. The latter established the textbook explanation of the phenomenon, based on conservation of mass and momentum, ignoring any additional effects such as the slope of the channel, the friction with the bottom and viscosity, or surface tension. Only during the past few decades these effects have begun to be incorporated, often via approaches that depart from that of B\'elanger~\cite{bowles1992,bohr1993,higuera1994,bohr1997,bonn2009,devita2020}.

A consequence of ignoring viscosity is that the super- and subcritical flow regimes do not connect continuously but have to be artificially linked to each other via vertical segments obeying the Rankine-Hugoniot shock relations~\cite{rayleigh1914}. Until today, this remains the standard way of describing hydraulic jumps~\cite{mejean2017,dhar2019}. A major drawback of this approach, however, is that the shock structure of the jump is unaccounted for. We elucidate this structure by analyzing the \textit{viscous} Saint-Venant equations~\cite{needham1984,merkin1986,kranenburg1992} and we derive an analytic expression for the jump length $L$, thereby settling a long standing issue in laminar channel flow.

The forces governing the shape of the hydraulic jump are gravity, surface tension, and effects due to viscosity. Just as for usual surface waves, gravity is dominant at large scales while surface tension becomes important at small scales. Viscous effects (especially the normal stresses) are instrumental in safeguarding the continuity of the fluid profile. The latter come into play in the flow regions where the height and velocity change rapidly, being the physical mechanism that prevents the profile from becoming infinitely steep. 

The familiar jump in the kitchen sink is on the relatively small scale where gravity, surface tension, and viscosity all contribute, and is therefore despite its commonness a surprisingly intricate case~\cite{duchesne2019,bhagat2018}. On the largest scale, one finds the fully developed turbulent jumps of spillways and rapids, where gravity takes the undisputed lead. The present paper deals with jumps on the intermediate scale where surface tension becomes insignificant, leaving gravitational and viscous effects as the two main factors. Such jumps may be encountered in irrigation ditches or tilted channels on the laboratory scale under laminar or smoothly turbulent flow conditions. The fluid in question may be taken to be water, but the description will become increasingly accurate if one uses a Newtonian fluid with a higher viscosity and even lower surface tension, e.g. silicon oil or castor oil. The Reynolds number can then be kept effortlessly at the moderate levels of laminar or just turbulent flow, while the capillary effects of surface tension (already very minor at this scale) can safely be ignored. As for the Froude number of the incoming flow, this must naturally exceed $1$ yet it should remain bounded within the realm of what is traditionally known as a ``weak jump'', for which the heights of the incoming and outgoing flows are of the same order.  

\section{The Saint-Venant equations }\label{sec:Saint-Venant}

Shallow water flow in a channel is aptly described by the height $h(x,t)$ of the flowing sheet and its depth averaged velocity $\bar{u}(x,t)$. This one-dimensional description does not account for the velocity component in the vertical direction, which obviously \textit{must} exist in the jump region, since the fluid has to rise to a larger height here. Our analysis should therefore be considered as a mean field theory, giving the height profile $h(x,t)$ without any special effects (such as undulations, surface rolls or eddies) arising from the internal dynamics in the fluid sheet.

The two quantities $h(x,t)$ and $\bar{u}(x,t)$ are governed by the Saint-Venant equations, i.e., the two coupled partial differential equations (PDEs) expressing respectively the mass balance
\begin{equation}\label{massbal}
h_t + (h \bar{u})_x = 0,
\end{equation}
\noindent and the momentum balance
\begin{equation} \label{mombal}
(h \bar{u})_t + (h \bar{u}^2)_x = gh \sin \zeta - ( \frac{1}{2} g h^2\cos \zeta )_x - C_f \bar{u}^2 + \nu (h \bar{u}_x)_x . 
\end{equation}

The terms on the right hand side of Eq.~(\ref{mombal}) represent the various forces acting on the water sheet. In order of appearance: (i) The gravity component in the $x$-direction, where $g=9.81~\textrm{m}/\textrm{s}^{2}$ is the gravitational acceleration and $\zeta$ the inclination angle of the channel. (ii) The pressure gradient arising from the variations in $h(x,t)$. (iii) The friction with the bottom of the channel arising from the viscous shear stresses. The form we use is known as Ch\'ezy's formula, valid in the laminar and smoothly turbulent regimes. (iv) The force due to the viscous normal stresses, where $\nu$ denotes an empirical positive coefficient with the dimensions of a kinematic viscosity. The latter term was absent from the original Saint-Venant equations. Several forms, suitable for different flow regimes, have been proposed over the years; see e.g. Needham and Merkin (1984) \cite{needham1984}. The one adopted here was derived by Kranenburg (1992) \cite{kranenburg1992} for laminar and moderately turbulent flows. Typically, since the flow is shallow, the bottom friction is the primary source of dissipation. So we take $\nu$ to be small and consequently, our results do not depend substantially on the precise form of the normal viscous term. Finally, as discussed in the Introduction, the jumps we are interested in are too large to be noticeably affected by surface tension; therefore, any capillary force terms have been left out of Eq.~(\ref{mombal}).    
     
The above set of equations has been used successfully to reproduce travelling waveforms in open channel flow, such as roll waves~\cite{needham1984,merkin1986,kranenburg1992,balmforth2004}. Here we employ the same equations to describe a \textit{stationary} waveform, namely the standing hydraulic jump. In order to capture this type of wave, we seek solutions of the form $h=h(x)$ and $\bar{u}=\bar{u}(x)$. This implies that the derivatives with respect to $t$ in Eqs.~(\ref{massbal}) and (\ref{mombal}) vanish identically. 

The time-independent version of the mass balance Eq.~(\ref{massbal}) can readily be integrated, yielding $h \bar{u} = Q$, where $Q$ is the constant flux of water (per unit width) in the channel. Inserting the relation $\bar{u}(x) = Q h(x)^{-1}$ into the time-independent version of Eq.~(\ref{mombal}), we obtain a second-order ordinary differential equation (ODE) for $h(x)$:
\begin{equation}\label{viscousmotheq}
\frac{\nu}{Q} h h'' = \frac{\nu}{Q} (h')^2 + \left( 1 - \frac{h^3}{h_c^3} \right) h' -C_f \left( 1 - \frac{h^3}{h_n^3} \right), 
\end{equation}
\noindent where the prime stands for differentiation with respect to $x$, the combination $Q/ \nu$ is the effective Reynolds number of the flow (since $\nu$ is an effective viscosity coefficient), while the heights $h_c$ and $h_n$ are given by
\begin{equation}\label{hchn}
h_c = \left( \frac{Q^2}{g \cos \zeta} \right)^{1/3}, ~~~~ h_n = \left( \frac{C_f Q^2}{g \sin \zeta} \right)^{1/3}.
\end{equation}
\noindent In the context of gradually varied flow, these are known as the \textit{critical} and \textit{natural} height, respectively~\cite{bakhmeteff1932,rouse1946}. At $h=h_c$ the Froude number $F = \bar{u}/\sqrt{gh \cos \zeta} = (h_c/h)^{3/2}$ equals 1, marking the border between supercritical flow ($F >1$, $h<h_c$) and subcritical flow ($F<1$, $h>h_c$). The height $h_n$ is the water thickness under uniform flow conditions, i.e., when $h$ and $\bar{u}$ are constant. It follows from Eq.~(\ref{mombal}) by setting all derivatives equal to zero, leaving only the forces of gravity and friction to balance each other.

The quotient $(h_c/h_n)^3 = \tan \zeta / C_f$ forms the basis for the standard classification of channels into two categories: (1) Those that are hydraulically \textit{mild} for $h_c < h_n$, and (2) those that are hydraulically \textit{steep} for $h_c > h_n$. Equivalently, one might talk about hydraulically ``rough'' and ``smooth'' channels, respectively. 

Equation~(\ref{viscousmotheq}) shows precisely what the inclusion of viscosity has added to the description of the hydraulic jump, namely, the terms with $h h''$ and $(h')^2$. Indeed, in the inviscid limit $\nu \rightarrow 0$, Eq.~(\ref{viscousmotheq}) reduces to the relation that is traditionally used to describe gradually varied flow, known as the `backwater equation'~\cite{chanson2009}: 
\begin{equation}\label{inviscidmotheq}
h' = C_f \frac{ 1 - (h/h_n)^3}{1 - (h/h_c)^3} = \tan \zeta ~\frac{h^3-h_n^3}{h^3-h_c^3}. 
\end{equation}

Until now, the standard procedure was to derive from this inviscid relation the behaviors upstream and downstream of the jump, and to patch these together. For mild channels, there are three distinct profiles: $M_1$ (for $h>h_n$), $M_2$ ($h_c < h < h_n$), and $M_3$ (for $h < h_c$), see Fig.~\ref{fig2}a for their viscous counterparts. Of these three, the profile $M_3$ is the only supercritical one, so the jump connects $M_3$ either to $M_2$ or to $M_1$. Also for steep channels there are three distinct profiles: $S_1$ (for $h>h_c$), $S_2$ ($h_n < h < h_c$), and $S_3$ (for $h < h_n$), cf. Fig.~\ref{fig2}b. In this case, both $S_2$ and $S_3$ are supercritical, hence the jump connects either of these profiles to $S_1$. Any of these jumps links a region of supercritical flow to a subcritical one, and hence occurs when the level of the water passes through the critical height $h_c$. In the inviscid expression Eq.~(\ref{inviscidmotheq}), the slope $h'$ becomes infinite for $h=h_c$, so the jump appears as a discontinuity in the profile $h(x)$, i.e., as a vertical segment connecting the super- and subcritical regimes. This unphysical feature is cured, as we will demonstrate, by the inclusion of viscosity. The crux of the matter is that the solutions of the viscous Eq.~(\ref{viscousmotheq}) cross the height $h_c$ with a \textit{finite} slope.

\section{Dynamical Systems approach}\label{sec:dynamicalsystem}

The second-order ODE Eq.~(\ref{viscousmotheq}) can be cast in the form of a Dynamical System consisting of two first-order ODEs, as follows (with $R = Q/ \nu$ being the effective Reynolds number):
\begin{subequations}\label{DS}
\begin{eqnarray}
h' ~&=&~ s ~=~ f_1(s), \\  
s' ~&=&~ \frac{s^2}{h} + R \left( 1 - \frac{h^3}{h_c^3} \right) \frac{s}{h}~-C_f R \left( 1 - \frac{h^3}{h_n^3} \right) \frac{1}{h} ~= f_2(h,s). 
\end{eqnarray}
\end{subequations}
\noindent We have chosen to denote $h'$ by the letter $s$ since this quantity represents the \textit{slope} of the water sheet. Note that the division by $h$ in Eq.~(\ref{DS}b) poses no problem, since the flow thickness can never drop to zero, or else the velocity $\bar{u}$ would have to become infinite in order to maintain the constant flux $Q$.

\begin{figure*}[!]
\begin{center}
\includegraphics[width=0.62\textwidth]{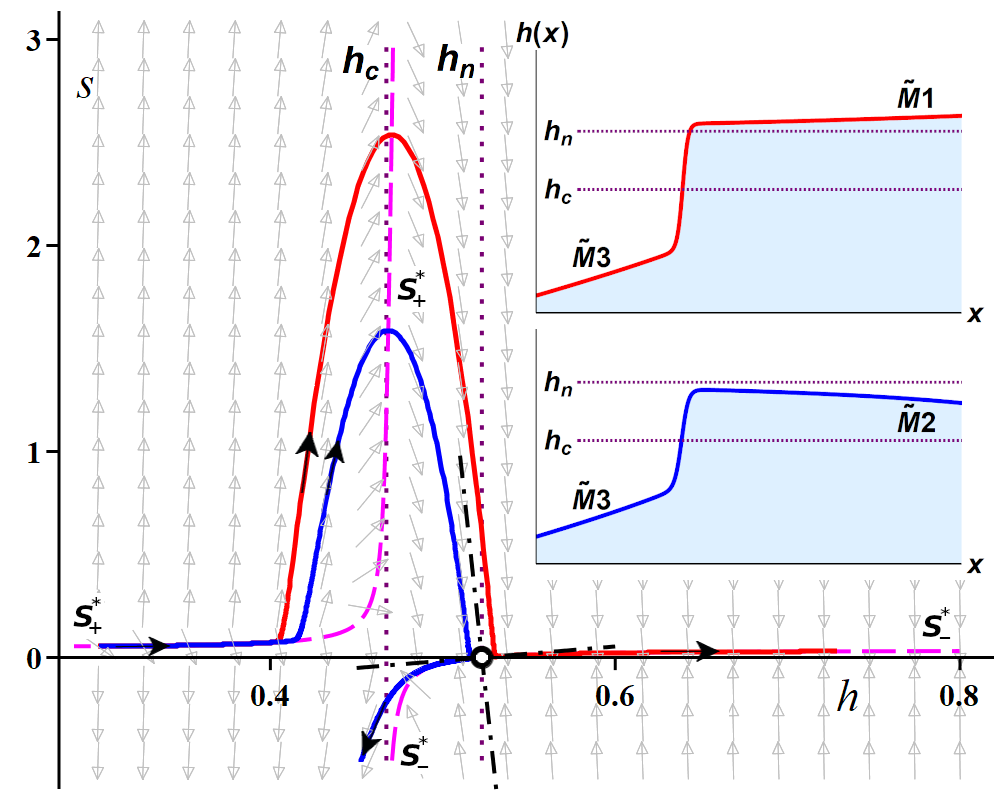}
\\~\\
\includegraphics[width=0.62\textwidth]{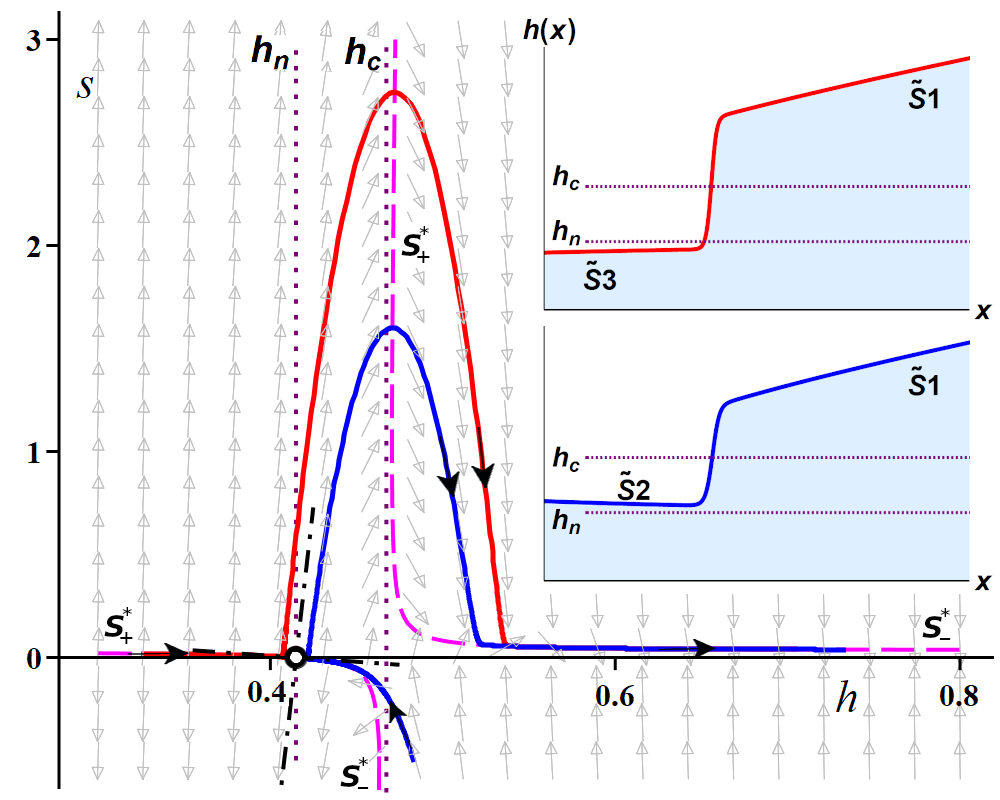}
\caption{Phase portraits of the dynamical system (\ref{DS}a)-(\ref{DS}b) for mild (a) and steep (b) channels, which are seen to be each other's mirror image. The solid red and blue curves represent the two types of hydraulic jumps occurring in each channel; the jump region always manifests itself as a near-parabolic orbit. (a) Mild: The curves travel together in the \textit{supercritical} regime, and then go their separate ways; the associated profiles $h(x)$ (see insets) are the jumps $\widetilde{M}3 \rightarrow \widetilde{M}1$ and $\widetilde{M}3 \rightarrow \widetilde{M}2$. (b) Steep: The curves start out differently, coming together in the \textit{subcritical} regime; the profiles $h(x)$ (insets) are the jumps $\widetilde{S}3 \rightarrow \widetilde{S}1$ and $\widetilde{S}2 \rightarrow \widetilde{S}1$. The dotted lines indicate the levels $h_c$ and $h_n$, and the grey arrows in the background show the direction field. Parameter values: $\zeta = 2^\circ$, $\nu =0.01 \textrm{m}^2/\textrm{s}$, $Q=1 \textrm{m}^2/\textrm{s}$, with $C_f = 1.4 \tan \zeta$ (mild) and $C_f = 0.7 \tan \zeta$ (steep).} \label{fig2}
\end{center}
\end{figure*}

The fixed points of the system (\ref{DS}a)-(\ref{DS}b) are found by setting $h'=0$ and $s'=0$ simultaneously. From (\ref{DS}a) we see that every fixed point must necessarily have $s=0$, corresponding to a flat water sheet, and substituting this in Eq.~(\ref{DS}b) one obtains $h^3 = h_n^3$. Therefore, the system has one real fixed point $(h,s) = (h_n,0)$; the other two roots of $h^3 = h_n^3$ are complex conjugate and of no physical relevance. The linear stability analysis around $(h_n,0)$ reveals that the fixed point is a saddle, with two real eigenvalues of opposite sign.

A key role in the dynamics of the system is played by the \textit{nullclines} given by $f_1(s)=0$ and $f_2(h,s)=0$, respectively. The first of these is simply the horizontal axis $s=0$. The second one has a more intricate form:
\begin{equation}\label{nullcline}
s_{\pm}^{*}(h) = \frac{R}{2 h_c^3} ~ \{ ~ h^3 -h_c^3 ~\pm \sqrt{\left( h^3 -h_c^3 \right)^2 - \frac{4 C_f h_c^6 }{R h_n^3}(h^3-h_n^3)} ~\}, 
\end{equation}
\noindent shown in Figs.~\ref{fig2}a,b as a dashed line. Evidently, the fixed point $(h_n,0)$ lies at the intersection of the two nullclines.    

The nullcline $s_{\pm}^{*}(h)$ consists of two branches. For mild channels, these branches intersect the critical line $h=h_c$. For steep channels on the other hand, for which $h_n < h_c$, the expression under the square root in Eq.~(\ref{nullcline}) becomes negative around $h = h_c$ and hence the branches leave a gap there, as seen in Fig.~\ref{fig2}b.

\begin{figure}
\begin{center}
\includegraphics[width=0.7\textwidth]{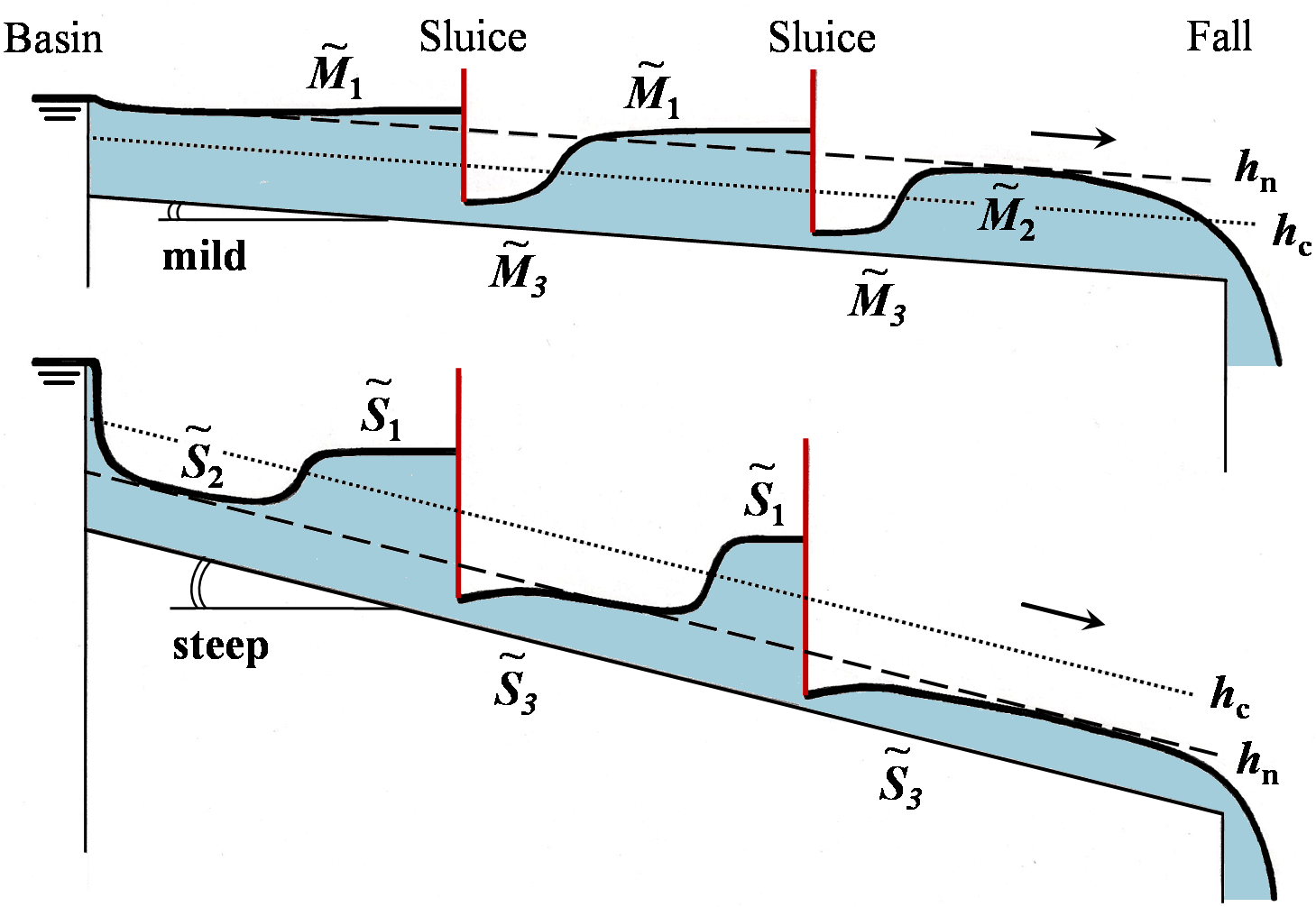}
\caption{Overview of the four types of hydraulic jumps that are encountered in laminar or smoothly turbulent open channel flow, arising as stationary solutions of the viscous Saint-Venant equations (\ref{massbal})-(\ref{mombal}). (a) The jumps $\widetilde{M}3 \rightarrow \widetilde{M}1$ and $\widetilde{M}3 \rightarrow \widetilde{M}2$ (Fig.~\ref{fig2}a) occur in a channel with mild slope with the aid of two sluice gates. (b) In a steep channel, the jumps $\widetilde{S}2 \rightarrow \widetilde{S}1$ and $\widetilde{S}3 \rightarrow \widetilde{S}1$ (Fig.~\ref{fig2}b) are materialized in a similar manner.} \label{fig3}
\end{center}
\end{figure}

Inside the jump region, the nullcline $s_{\pm}^{*}(h)$ intersects the phase space trajectory exactly at its maximum, corresponding to the inflection point of the jump (which lies close to the critical level $h=h_c$). Outside the jump region, on either side, the trajectories can be shown to converge to this nullcline, which thereby governs the system's asymptotic behavior, cf. Figs.~\ref{fig2}a,b. Specifically, for $h \rightarrow 0$ the slope of $h(x)$ is     
\begin{equation}\label{slopeAtZero}
\lim_{h \rightarrow 0} s ~=~ \small{\frac{R}{2}} \left( \sqrt{1+4 \frac{C_f}{R}} ~-~ 1 \right) =~ C_f - \frac{C_f^2 \nu }{Q} + O(\nu^2),
\end{equation}
\noindent which is the viscous correction to the classical result for the asymptotic slopes of the profiles $M3$ and $S3$. At the other end of the spectrum, for $h \rightarrow \infty$, we find $s = \tan \zeta$, corresponding to a horizontal water profile with respect to the laboratory~\cite{rouse1946}. This situation is typically encountered before a sluice gate (see Fig.~\ref{fig3}).   

Figure~\ref{fig2}a gives a detailed view of the viscous jump in \textit{mild} channels. The system parameters do not mimic any specific experiment, but have rather been chosen in such a way as to illustrate the structure of the phenomenon as clearly as possible. The phase portrait shows marked similarities with that presented by Bohr \textit{et al.}~\cite{bohr1997}, who already in 1997 applied a dynamical systems approach to the study of hydraulic jumps. For small values of $h$, where the flow is supercritical, the trajectories run close to the nullcline $s_{+}^{*}(h)$ of Eq.~(\ref{nullcline}). Then they depart from it, along a nearly parabolic orbit that follows the stable manifold of the saddle point. They cross the critical value $h_c$ close to the top of the orbit --where the flow becomes subcritical-- and descend to the close neighbourhood of the saddle $(h_n,0)$. Here the trajectories either go to the right, to the left, or (in the borderline case) hit exactly upon the saddle . The red trajectory in Fig.~\ref{fig2}a reconnects to the rightmost branch $s_{-}^{*}(h)$ of the nullcline, forming the profile $\widetilde{M}1$, which eventually attains the constant slope $s = \tan \zeta$. The tilde is used to distinguish the viscous profile from its inviscid counterpart of the classic theory~\cite{rouse1946}.  

The blue trajectory, on the other hand, bends off to the left of the saddle, i.e., to smaller values of $h$ with negative slope $s$. It drops to the critical level $h_c$ and even goes below it, thereby rendering the flow supercritical again. This goes beyond the inviscid result according to which $M2$ always remains purely subcritical, and permits the flow to discharge at supercritical conditions as observed in practice~\cite{chow1973}. The supercriticality cannot be pushed too far, however, because at some point the height starts falling sharply towards zero (and the velocity $\bar{u}=Q/h$ diverges), meaning that our analysis breaks down. So the blue trajectory in Fig.~\ref{fig2}a loses its significance at some certain level below $h=h_c$. For the present study this does not matter though, since the jump ($\widetilde{M}3 \rightarrow \widetilde{M}2$) is contained in the preceding parts of the trajectory.

The jump trajectories and profiles for a \textit{steep} channel are depicted in Fig.~\ref{fig2}b. The red and blue trajectories now start out from \textit{different} levels, but both approach the vicinity of the saddle point $(h_n,0)$, from which they escape along near-parabolic orbits that are organized around the saddle's unstable manifold. As before, they cross the critical level $h=h_c$ close to their maximum and when they come down again, both trajectories converge to the rightmost branch of the nullcline $s_{-}^{*}(h)$, which represents the familiar profile with slope $s = \tan \zeta$. In analogy with the terminology from the inviscid theory, we call this the $\widetilde{S}1$ profile.

It is apparent from the phase space trajectories in Figs.~\ref{fig2}a,b that the jumps for steep channels are \textit{mirror images} of those for mild ones, illustrating the profound relation between these two classes of hydraulic jumps.
  
The above findings are recapitulated in Fig.~\ref{fig3}, presenting an overview of all four jump types in a setup where they might be observed experimentally, involving channels with two sluice gates and an outlet (or ``fall''). The figure is an updated and extended version of a classic textbook picture [\cite{rouse1946}, p. 228], this time with all types of jumps included. Importantly, thanks to viscosity, the jumps in the present description are \textit{continuous}, whereas in the inviscid theory they were merely vertical segments.
  
At this point, we note that the question of which jump will be realized in practice is dictated by the boundary conditions. In the case of a hydraulically mild channel, the \textit{downstream} conditions decide the issue: confronted with a sluice gate, $\widetilde{M}3$ will necessarily jump to the $\widetilde{M}1$ branch, whereas before a fall it has to take the $\widetilde{M}2$ branch. For a hydraulically steep channel, by contrast, the flow is controlled by the \textit{upstream} conditions: any jump now has to end in $\widetilde{S}1$, since this is the only subcritical branch, and whether it will do so from $\widetilde{S}2$ or $\widetilde{S}3$ depends on whether the height from which the flow starts is above or below $h_n$, respectively.

This difference between downstream and upstream control beautifully corroborates the mirror symmetry that exists between the jumps in mild and steep channels. The physical reason for this can be traced back to the fact that information in the supercritical regime ($F>1$) can only travel downstream, because the fluid travels faster than any surface wave; hence the choice between $\widetilde{S}2$ or $\widetilde{S}3$ must necessarily be decided at the upstream end of the flow sector in question. In the subcritical regime ($F < 1$) information can travel in both directions, which means that the choice between $\widetilde{M}1$ and $\widetilde{M}2$ can be made at the downstream end of the flow sector~\cite{rouse1946}.     

\section{Length of the jump}\label{sec:legth}

\begin{figure}
\begin{center}
\includegraphics[width=0.7\textwidth]{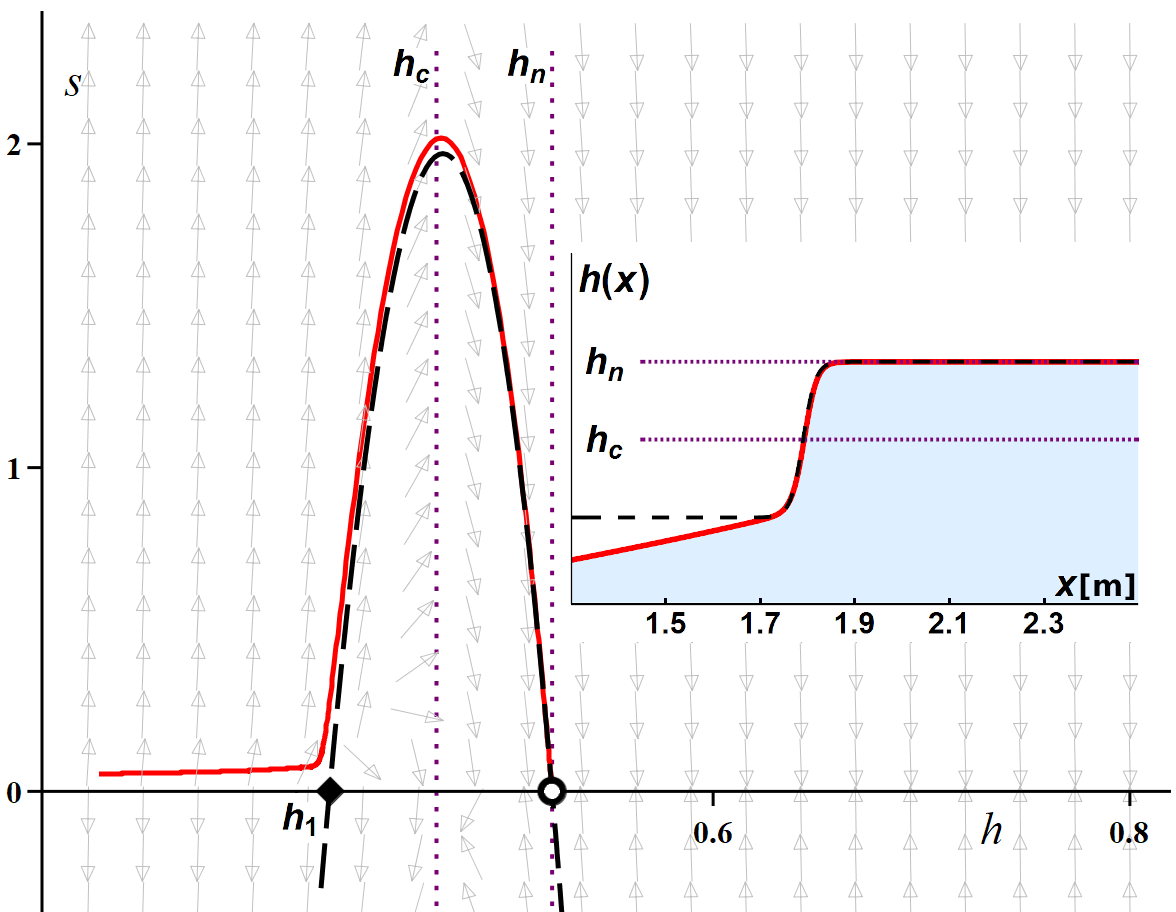}
\caption{Phase space trajectory (solid red curve) in a hydraulically mild channel of the borderline jump ending precisely at $h=h_n$, being the stable manifold of the saddle point $(h_n,0)$. The black dashed curve denotes the cubic expression for $s(=h')$ given by Eq.~(\ref{length4bis}), which is seen to be an excellent approximation in the jump region $h_1 < h < h_n$. This close agreement is used to derive the analytical expression Eq.~(\ref{lengthL2}) for the jump length $L$. Inset: the tanh-profile Eq.~(\ref{hprofile}) corresponding to the approximation. The parameter values are the same as in Fig.~\ref{fig2}a.} \label{fig4}
\end{center}
\end{figure}

For slightly inclined channels, with $\tan \zeta \approx C_f$, Eq.~(\ref{hchn}) shows that $h_n$ and $h_c$ are indeed of the same order, in accordance with our assumption of ``weak jumps''. The expressions $1 - (h/h_c)^3$ and $1 - (h/h_n)^3$ appearing in Eq.~(\ref{viscousmotheq}) are then of the same order as well and, given the fact that in the jump region the slope $h' \gg \tan \zeta \approx C_f$, the last term in Eq.~(\ref{viscousmotheq}) may safely be neglected. Thus, in the jump region, Eq.~(\ref{viscousmotheq}) is well approximated by (with $Q/ \nu = R$ as usual):
\begin{equation}\label{length1}
h h'' \approx (h')^2 + R \left( 1 - (h/h_c)^3 \right) h'. 
\end{equation}
\noindent Rewriting this second-order ODE as follows:  
\begin{equation}\label{length2}
\frac{h'' h - h' h'}{h^2} =  R \left( h^{-2}h' - \frac{1}{h_c^3} h h' \right), 
\end{equation}
\noindent it is readily integrated to yield    
\begin{equation}\label{length3}
\frac{h'}{h}  =  -R \left( h^{-1} + \frac{1}{2h_c^3} h^2 \right) + B, 
\end{equation}
\noindent with the integration constant $B$ being determined by the fact that the jump trajectory passes close by the saddle point $(h,h') = (h_n,0)$, i.e., $B = R(h_n^{-1} + h_n^2/(2h_c^3))$. This leaves us with the following first-order ODE (cf. Fig.~\ref{fig4}):
\begin{equation}\label{length4bis}
h' =  -Ah^3 + Bh -R = A (h_n-h)(h-h_1)(h-h_2), 
\end{equation}

\noindent where $A= R/(2h_c^3)$. In the last step we have written the cubic expression in terms of its roots, by first extracting the root $h=h_n$ associated with the saddle point $(h,h') = (h_n,0)$. The other two roots $h_1, h_2$ are not associated with any fixed points of the full dynamical system but arise from the approximation made in Eq.~(\ref{length1}). They are real-valued and of opposite sign. Indeed, by noticing that $R= -Ah_n h_1 h_2$, one has $h_1 = -2h_c^3/(h_n h_2)$. For a typical jump (roughly symmetric around $h_c$) one finds $h_1 \approx 2h_c - h_n$ and $h_2 \approx -2h_c^3/[h_n (2h_c - h_n)]$; since $h_1$ and $h_n$ are both of the order of $h_c$, one has $h_2 \approx -2h_c$.

Equation~(\ref{length4bis}) is a separable ODE. It may be cast in the form $dx=dh/[A (h_n-h)(h-h_1)(h-h_2)]$ and then be solved analytically by decomposing the right hand side in its partial fractions. The result is an elaborate expression involving logarithms. For our purposes, it is sufficient to focus upon the region $h_1 < h < h_n$, where $h \sim h_c$ and hence $h-h_2 \approx 3h_c$. In this region, Eq.~(\ref{length4bis}) is well approximated by $h' = 3h_cA (h_n-h)(h-h_1)$, which complies with the nearly parabolic shape of the jump trajectory in phase space; see Figs.~\ref{fig2} and \ref{fig4}. This reduced ODE upon integration yields:
\begin{equation}\label{hprofile}
h(x) \approx \small{\frac{1}{2}}(h_n+h_1) + {\small{\frac{1}{2}}}(h_n-h_1) \tanh \left( \alpha x \right) ,
\end{equation}
\noindent where $\alpha = \small{\frac{3}{2}} h_cA(h_n-h_1) = 3R(h_n-h_1)/(h_n+h_1)^2$. The integration constant here has been chosen such that the jump is centred around $x=0$, i.e. $h(0)=\small{\frac{1}{2}}(h_n+h_1) \approx h_c$. The approximate profile Eq.~(\ref{hprofile}) traverses the jump amplitude $h_n-h_1$ in a symmetrical fashion.

Now, the jump length $L$ may be defined as the distance in the $x$-direction in which $h(x)$ completes $99 \%$ of its course. Since $\tanh(\alpha x) = 0.99$ at $\alpha x = 2.65$, this gives $L = 2 \times 2.65/ \alpha = 1.76 (h_n+h_1)^2/[R(h_n-h_1)]$. With $h_n-h_1 \approx 2h_c(1-h_1/h_c)$ and recalling that the Froude number of the supercritical incoming flow can be expressed as $F_1=(h_c/h_1)^{3/2}$, we arrive at 
\begin{equation}
\label{lengthL2}
L ~=~ \frac{3.53~h_c}{R (1-F_1^{-2/3})}~,~~~~\textrm{with}~~  F_1 > 1.
\end{equation}
\noindent The jump length given by this expression is found to be in excellent agreement with the value obtained from numerically solving the full dynamical system (\ref{DS}a)-(\ref{DS}b). For example, for the jump depicted in Fig.~\ref{fig4}, with $h_c = 0.467$~m, $R = 100$ and $F_1=1.21$, both Eq.~(\ref{lengthL2}) and the numerical solution give $L = 0.14$~m.  

Equation~(\ref{lengthL2}) reflects the fact that the jumps considered here result from an interplay between \textit{gravity} and \textit{viscosity}, represented by the Froude number $F_1$ and the Reynolds number $R$, respectively. $L$ is inversely proportional to $R$, i.e., it grows linearly with $\nu$, which stands to reason given the flattening effect of the viscous forces. In the limit $\nu \rightarrow 0$ the length vanishes, reproducing the infinitely steep jump of the classical inviscid analysis. As for the Froude number, the length $L$ is maximal in the limit $F_1 \rightarrow 1$ and decreases monotonically for growing $F_1$. For $F_1 \gg 1$, the length becomes independent of $F_1$. 

\begin{figure}
\begin{center}
\includegraphics[width=0.7\textwidth]{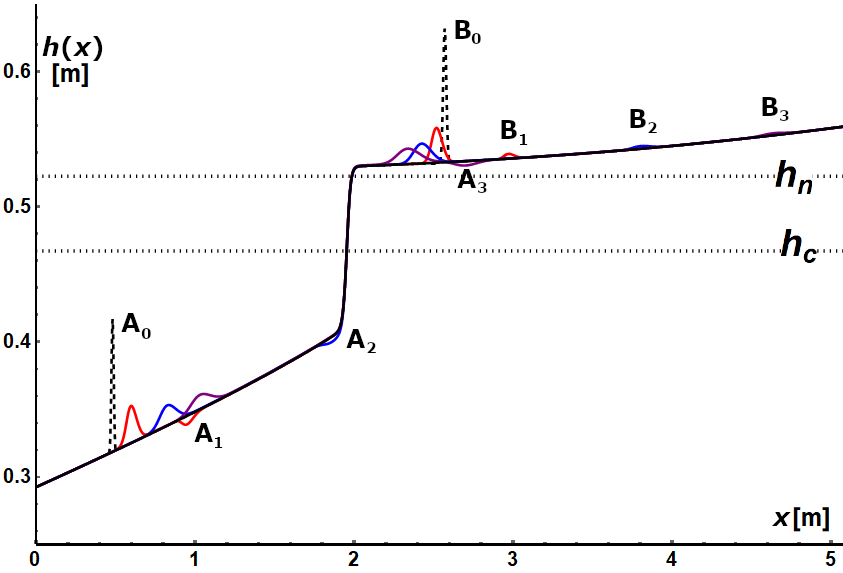}
\caption{Stability of the hydraulic jump $\widetilde{M}3 \rightarrow \widetilde{M}1$: two initial perturbations $\textrm{A}_0$ and $\textrm{B}_0$ are positioned on the jump's lower and upper branch, and their subsequent evolution is computed from the viscous Saint-Venant equations (\ref{massbal})-(\ref{mombal}). Each perturbation breaks into two wave packets that decay with time, evidencing the stability of the jump. As a guide to the eye, three successive snapshots of the rightmost wave packet of each perturbation are indicated by $\textrm{A}_i$ and $\textrm{B}_i$, $i=1,2,3$. On the lower branch the packets propagate downstream since $F>1$; they pass through the critical level $h=h_c$ and enter the subcritical regime. The packets generated on the upper branch travel in both directions (because here $F<1$), with the one moving upstream never getting past $h = h_c$. } \label{fig5}
\end{center}
\end{figure}

\section{Conclusion}\label{sec:conclusion}

The hydraulic jumps in laminar or mildly turbulent open channel flow are fully captured, complete with their mean-field shock structure, as stationary solutions of the viscous Saint-Venant equations. In phase space, the jumps manifest themselves as trajectories that leap from the supercritical branch of the nullcline to the opposite subcritical branch via a pronounced, approximately parabolic orbit. This orbit follows the stable/unstable manifold of the saddle point $(h,h') = (h_n,0)$, for mild/steep channels respectively, while the nullcline $s_{\pm}^{*}(h)$ given by Eq.~(\ref{nullcline}) provides the exact form of the gradually varied profiles before and after the jump.  

The important issue of the stability of the thus obtained profiles can be settled by inserting them, in perturbed form, as initial data into the governing PDEs (\ref{massbal})-(\ref{mombal}). An example is shown in Fig.~\ref{fig5}, where the perturbations are seen to decay with time, confirming the stability of the jump. Interestingly, the perturbation on the subcritical upper branch is seen to send small waves in both directions (because here $F<1$), whereas the perturbation on the supercritical lower branch can make itself be felt only in the downstream direction. See also our discussion about the upstream and downstream control in the context of Fig.~\ref{fig3}. This one-way propagation of information in the supercritical regime has given rise to an intriguing analogy between hydraulic jumps and white holes in cosmology~\cite{volovik2005,jannes2011}.  

The phase space representation has yielded an unprecedented geometrical insight into the structure of the jumps. Among many other things, it has brought to light the mirror symmetry that exists between the jumps in mild and steep channels, see Fig.~\ref{fig2} and the overview in Fig.~\ref{fig3}, where the two possible jump types for mild channels are presented along with their two counterparts for steep channels.

Finally, the Dynamical Systems description adopted in the present paper has enabled us to derive the analytic approximative expression (\ref{lengthL2}) for one of the most prominent features of the hydraulic jump, namely, its length.


\begin{thebibliography}{99}

\bibitem{bakhmeteff1932}
B.A. Bakhmeteff, \textit{Hydraulics of open channels} (McGraw-Hill, New York, 1932).

\bibitem{chanson2009}
H. Chanson, J. Hydraul. Eng. {\bf 135}, 159 (2009).

\bibitem{bowles1992}
R.I. Bowles and F.T. Smith, J. Fluid Mech. \textbf{242}, 145 (1992).

\bibitem{bohr1993}
T. Bohr, P. Dimon, and V. Putkaradze, J. Fluid Mech. \textbf{254}, 635 (1993).

\bibitem{higuera1994}
F.J. Higuera, J. Fluid Mech. \textbf{274}, 69 (1994).

\bibitem{bohr1997}
T. Bohr, V. Putkaradze, and S. Watanabe, Phys. Rev. Lett. \textbf{79}, 1038 (1997).

\bibitem{bonn2009}
D. Bonn, A. Andersen, and T. Bohr, J. Fluid Mech. \textbf{618}, 71 (2009).

\bibitem{devita2020}
F. De Vita, P.-Y. Lagr\'ee, S. Chibbaro, S. Popinet, Eur. J. Phys./B Fluids \textbf{97}, 233 (2020).

\bibitem{rayleigh1914}
Lord Rayleigh, Proc. R. Soc. London A \textbf{90}, 324 (1914).
%
\bibitem{mejean2017}
S. Mejean, T. Faug, and I. Einav, J. Fluid Mech. {\bf 816}, 331 (2017).
%
\bibitem{dhar2019}
M. Dhar, G. Das, and P.K. Das, J. Fluid Mech. \textbf{884}, A11 (2019).

\bibitem{needham1984}
D.J. Needham and H.J. Merkin, Proc. R. Soc. London A \textbf{394}, 259 (1984).

\bibitem{merkin1986}
H.J. Merkin and D.J. Needham, Proc. R. Soc. London A \textbf{405}, 103 (1986).

\bibitem{kranenburg1992}
C. Kranenburg, J. Fluid Mech. \textbf{245}, 249 (1992).

\bibitem{duchesne2019}
A. Duchesne, A. Andersen, and T. Bohr, Phys. Rev. Fluids \textbf{4}, 084001 (2019)
%
\bibitem{bhagat2018}
R.K. Bhagat, N.K. Jha, P.F. Linden, and D.I. Wilson, J. Fluid Mech. \textbf{851}, R5 (2018).

\bibitem{balmforth2004}
N.J. Balmforth and S. Mandre, J. Fluid Mech. \textbf{514}, 1 (2004).

\bibitem{rouse1946}
H. Rouse, \textit{Elementary Mechanics of Fluids} (Dover, 1946).

\bibitem{chow1973}
V.T. Chow, \textit{Open Channel Hydraulics} (McGraw-Hill, New York, 1973

\bibitem{volovik2005}
G. E. Volovik, JETP Lett. \textbf{82}, 624 (2005)

\bibitem{jannes2011}
G. Jannes, R. Piquet, P. Maïssa, C. Mathis and G. Rousseaux, Phys. Rev. E \textbf{83}, 056312 (2011)


\end{thebibliography}
\end{document}